\documentclass[sigconf,natbib]{acmart}

\PassOptionsToPackage{dvipsnames}{xcolor}

\usepackage{acronym}
\usepackage[noend]{algorithmic}
\usepackage{algorithm}
\usepackage{amsmath}
\usepackage{amssymb}
\usepackage{amsthm}
\usepackage{bbm}
\usepackage{beramono}
\usepackage{bm}
\usepackage{booktabs}
\usepackage[skip=0pt]{caption}
\usepackage{cleveref}
\usepackage[T1]{fontenc}
\usepackage{graphicx}
\usepackage{hyperref}
\usepackage{listings}
\usepackage{multirow}
\usepackage{natbib}
\usepackage[np, autolanguage]{numprint}
\usepackage{paralist}
\usepackage{siunitx}
\usepackage{soul}
\usepackage{subcaption}
\usepackage{tabularx}
\usepackage[dvipsnames]{xcolor}
\usepackage{enumerate}
\usepackage{enumitem}

\copyrightyear{2019}
\acmYear{2019}
\setcopyright{rightsretained}
\acmConference[SIGIR 2019]{SIGIR 2019}{July 21--25, 2019}{Paris, France}
\acmPrice{}
\definecolor{rj}{RGB}{0, 150, 0}
\definecolor{mdr}{RGB}{200, 0, 0}
\definecolor{ho}{RGB}{0, 50, 150}

\newcommand{\myparagraph}[1]{\smallskip\emph{#1.}}

\acrodef{IR}{Information Retrieval}
\acrodef{LTR}{Learning to Rank}
\acrodef{OLTR}{Online Learning to Rank}
\acrodef{CLTR}{Counterfactual Learning to Rank}
\acrodef{PL}{Plackett-Luce}
\acrodef{SGD}{Stochastic Gradient Descent}
\acrodef{IPS}{Inverse Propensity Scoring}
\acrodef{MSE}{Mean Squared Error}
\acrodef{DBGD}{Dueling Bandit Gradient Descent}
\acrodef{PDGD}{Pairwise Differentiable Gradient Descent}

\acrodef{CF-RANK}{Counterfactual SVMRank}
\acrodef{CF-DCG}{Counterfactual DCGRank}

\theoremstyle{definition}

\DeclareMathOperator*{\argmin}{argmin}

\theoremstyle{definition}

\allowdisplaybreaks

\author{Rolf Jagerman}
\orcid{0000-0002-5169-495X}
\affiliation{%
 \institution{University of Amsterdam}
 \city{Amsterdam}
 \country{The Netherlands}
}
\email{rolf.jagerman@uva.nl}

\author{Harrie Oosterhuis}
\affiliation{%
	\institution{University of Amsterdam}
	\city{Amsterdam}
	\country{The Netherlands}
}
\email{oosterhuis@uva.nl}
 
\author{Maarten de Rijke}
\orcid{0000-0002-1086-0202}
\affiliation{
 \institution{University of Amsterdam}
 \city{Amsterdam}
 \country{The Netherlands}
}
\email{derijke@uva.nl}

\title[Learning to Rank from User Interactions]{Learning to Rank from User Interactions:\\Comparing Counterfactual and Online Approaches}

\title[Learning to Rank from User Interactions]{To Model or to Intervene: A Comparison of Counterfactual and Online Learning to Rank from User Interactions}


\begin{document}

\copyrightyear{2019} 
\acmYear{2019} 
\setcopyright{acmlicensed}
\acmConference[SIGIR '19]{Proceedings of the 42nd International ACM SIGIR Conference on Research and Development in Information Retrieval}{July 21--25, 2019}{Paris, France}
\acmBooktitle{Proceedings of the 42nd International ACM SIGIR Conference on Research and Development in Information Retrieval (SIGIR '19), July 21--25, 2019, Paris, France}
\acmPrice{15.00}
\acmDOI{10.1145/3331184.3331269}
\acmISBN{978-1-4503-6172-9/19/07}



\begin{abstract}
\acf{LTR} from user interactions is challenging as user feedback often contains high levels of bias and noise.
At the moment, two methodologies for dealing with bias prevail in the field of \ac{LTR}:
\emph{counterfactual} methods that learn from historical data and model user behavior to deal with biases; and
\emph{online} methods that perform interventions to deal with bias but use no explicit user models.
For practitioners the decision between either methodology is very important because of its direct impact on end users.
Nevertheless, there has never been a direct comparison between these two approaches to unbiased \ac{LTR}.
In this study we provide the first benchmarking of both counterfactual and online \ac{LTR} methods under different experimental conditions.
Our results show that the choice between the methodologies is consequential and depends on the presence of selection bias, and the degree of position bias and interaction noise.
In settings with little bias or noise counterfactual methods can obtain the highest ranking performance; however, in other circumstances their optimization can be detrimental to the user experience.
Conversely, online methods are very robust to bias and noise but require control over the displayed rankings.
Our findings confirm and contradict existing expectations on the impact of model-based and intervention-based methods in \ac{LTR}, and allow practitioners to make an informed decision between the two methodologies.
\end{abstract}
\maketitle

\section{Introduction}
\label{sec:intro}

Interest in \acf{LTR} approaches that learn from user interactions has increased recently~\cite{joachims2017unbiased, bendersky2017learning, yue09:inter, hofmann_2013_reusing}.
Compared to learning from annotated datasets~\citep{liu2009learning}, implicit feedback obtained through user interactions matches user preferences  more closely~\cite{joachims2007evaluating}.
Furthermore, gathering interactions is much less costly than expert annotations~\cite{chapelle2011yahoo, liu2007letor}.
Additionally, unlike \ac{LTR} from annotated datasets, \ac{LTR} from user interactions can respect privacy-sensitive settings~\cite{bendersky2017learning}.
However, a big disadvantage of user interactions is that they often contain different types of bias and noise.
Hence, \ac{LTR} methods that learn from user interactions mainly focus on removing bias from the learning process~\cite{joachims2017unbiased, bendersky2017learning, Oosterhuis2018Unbiased}.

There are two main families of algorithms for \emph{unbiased} \ac{LTR} from user interactions: 
\begin{enumerate}[leftmargin=*]
	\item \ac{CLTR}~\cite{joachims2017unbiased}:
	These algorithms learn a ranking model from a historical interaction log, often collected using a production system. 
	They usually treat clicks as absolute relevance indicators and employ a form of re-weighing in order to debias interaction data. 
	Counterfactual methods have no experimental control; they avoid the risks associated with online interventions where untested rankings may be displayed. 
	A disadvantage is that they cannot explore and are limited to rankings displayed by the production system.

	\item \ac{OLTR}~\cite{yue09:inter}:
	This class of algorithms interactively optimize and update a ranking model after every interaction.
	They combat bias by \emph{interventions}, i.e., by displaying slightly modified rankings.
	This type of experimental control allows the learner to assess and learn novel rankings.
	Clearly, experimental control comes with a risk: untested rankings may hurt the user experience.

\end{enumerate}

\noindent%
For practitioners the decision whether to use counterfactual or online \ac{LTR} is important for practical deployment and user satisfaction with their ranking system.
E.g., if there are situations where \ac{CLTR} and \ac{OLTR} methods provide the same performance, the risks of interventions can be avoided.
However, if under some conditions \ac{CLTR} methods are unable to reach the same performance as online interventions promise to bring, an \ac{OLTR} method may be preferred.
Currently, there has not been a study comparing methods across the two methodologies.
As a result, 
it is currently unclear when which methods may be preferred, what benefits either methodology provides, and the scope of these benefits.
Direct comparisons between \ac{CLTR} and \ac{OLTR} are required to help advance the field of \ac{LTR} and inform its uptake.

A direct and fair comparison of counterfactual and online \ac{LTR} algorithms is non-trivial for several reasons.
First, \ac{CLTR} methods do not affect the user experience as they learn from historical data; in contrast, the user experience is a vital part of the evaluation of \ac{OLTR} methods.
Second, unlike \ac{OLTR} methods, \ac{CLTR} methods assume there is no selection bias, and proofs of their unbiasedness depend on this assumption.
Finally, the optimization problems for \ac{OLTR} and \ac{CLTR} methods are formulated differently -- therefore they may not be optimizing the same metrics and observed differences could be a consequence of this difference.

To the best of our knowledge, this is the first study to provide a direct comparison of \ac{CLTR} and \ac{OLTR} methods. 
Our main goal in this work is to answer the following question:
\begin{itemize}
\item[] \em How should \ac{LTR} practitioners choose which method to apply from either \emph{counterfactual} or \emph{online} \ac{LTR} methodologies?
\end{itemize}
In order to enable informed answers to this question, we address multiple aspects that are important to practitioners of both large-scale and small-scale \ac{LTR} systems.
First, we evaluate whether both approaches converge at the same level of performance, in other words, whether both approaches capture the true user preferences equally well.
Furthermore, we investigate how the learning outcomes are affected by different levels of selection bias, position bias and interaction noise.
Second, we evaluate how well the user experience is maintained during learning, since \ac{OLTR} methods could potentially deter users with inappropriate interventions. 
Thirdly, we investigate the effect of interventions by allowing counterfactual methods to execute periodic deployments; this simulates multiple steps of optimization and deployment as one would see in practice.

The research questions we address are:
\begin{enumerate}[align=left, label={\bf RQ\arabic*}, leftmargin=*]
	\item Do state-of-the-art counterfactual and online \ac{LTR} methods converge to the same level of performance? \label{rq:performance}
	\item Is the user experience the same for counterfactual methods as for online methods? \label{rq:userexperience}
	\item When do online interventions help the learning to rank algorithm? \label{rq:interventions}
\end{enumerate}

\noindent%
In this work we present the first direct comparison between \ac{CLTR} and \ac{OLTR} methods.
Our comparison leads to valuable insights as it reveals that, depending on the experimental conditions, a different methodology should be preferred.
In particular, our results show that \ac{OLTR} methods are more robust to selection bias, position bias and interaction noise.
However, under low levels of bias and noise \ac{CLTR} methods can obtain a significantly higher performance.
Furthermore, to our surprise we find that some properties asserted to pertain to \ac{CLTR} or \ac{OLTR} methods in previously published work appear to be lacking when tested.
For instance, in contrast with previously published expectations~\cite{Oosterhuis2018Unbiased} \ac{OLTR} is not substantially faster at learning than \ac{CLTR}, and while always assumed to be safe~\cite{wang2018position}, \ac{CLTR} may be detrimental to the user experience when deployed under high-levels of noise.

Our findings reveal areas where future \ac{LTR} work could make important advances, and moreover, allow practitioners to make an informed decision on which \ac{LTR} methodology to apply.


\section{Counterfactual Learning to Rank}
\label{sec:counterfactual}

\acf{CLTR}~\cite{joachims2017unbiased,agarwal2018counterfactual,ai2018unbiased-sigir} aims to learn a ranking model offline from historical interaction data.
Employing an offline approach has many benefits compared to an online one.
First, it is possible to try and iterate many different learning algorithms without needing to deploy them online.
Furthermore, it avoids the pitfalls and engineering overhead of having to deploy an online learning system.
Finally, models that are learned offline can be tested before actually being deployed online, alleviating some of the safety concerns surrounding \ac{OLTR}, such as the aggressive exploration of online methods that may place irrelevant items at high ranked positions~\cite{kveton2018bubblerank,wang2018position}.

A straightforward approach to \ac{LTR} from historical user interactions is to collect clicks and treat them as signals of relevance~\cite{joachims2017accurately}.
This is referred to as \emph{partial information feedback} because it only conveys information about the documents that the user \emph{has seen} and clicked on, but not other documents that the user \emph{could have} seen and clicked on. 
Traditional supervised learning algorithms expect data to be in a ``full information'' form, where it is exactly known which documents are relevant and which ones are not.
This is never the case in user interactions due to biases and noise.
As a solution, \ac{CLTR} provides a way to deal with partial information feedback.

\begin{table}
	\caption{Notation used throughout the paper.}
	\label{tbl:notation}
	\begin{tabular}{ll}
		\toprule
		\textbf{Notation} & \textbf{Description} \\
		\midrule
		$d$ & document \\
		$D$ & set of documents \\
		$R$ & ranked list \\
		$R_i$ & document placed at rank $i$ in ranked list $R$ \\
		$f_\theta(\cdot)$ & ranking model with parameters $\theta$ \\
		$c_i$ & 1 if document at rank $i$ was clicked, 0 otherwise \\
		$p_i$ & The propensity score at rank $i$ \\
		\bottomrule
	\end{tabular}
\end{table}

\subsection{Unbiased \ac{LTR} with biased feedback}
\citet{joachims2017unbiased} introduce a method to utilize interaction data in \ac{LTR}, by casting the problem as a counterfactual learning problem~\cite{swaminathan2015counterfactual}.
In~\cite{joachims2017unbiased}, it is assumed that the user does not examine all documents in a ranked list, and is more likely to observe documents at the top of the list than at the bottom; this is referred to as \emph{position bias}.
After a document is observed, a user will either judge it as relevant resulting in a click, or judge it as non-relevant.
More formally, the user observes document $d_i$ at rank $i$ with some probability $p_i$, called the \emph{propensity}.\footnote{The notation we use in the paper is listed in Table~\ref{tbl:notation}.}

If the propensity is known, it is possible to modify an existing learning algorithm and simply re-weigh interaction data according to the propensity scores using \acfi{IPS}.
\citet{joachims2017unbiased} take the SVMRank algorithm and modify it to optimize a re-weighted objective, where each click is re-weighted according to whether the click appeared at the top of the ranked list (thus with high propensity) or lower in the ranked list (thus with lower propensity).
Samples with high propensity are weighted less than samples with low propensity and vice versa.
We will assume the propensities are known a priori and discuss related work dealing with propensity estimation in Section~\ref{sec:counterfactual:propensities}

To formalize \ac{CLTR}, we consider a ranking function $f_{\text{production}}$ that produces a ranked list $R$ to be shown to the user in response to a query $q$.
When a user clicks on a document in this ranking, they are revealing to us that this document is relevant.
We denote a user's clicks by a 0/1 vector $c$:
$$
c_i = \left\{\begin{array}{ll}
1 & \text{if document }d_i\text{ was observed and judged relevant}, \\
0 & \text{otherwise}.
\end{array} \right.
$$
Note that it is possible for a user to click on more than one document during a session or click on no documents.
Since a user is more likely to observe top-ranked documents than lower-ranked ones, we are more likely to observe relevance signals of the top-ranked documents.
We denote the probability that a user observes the document at rank $i$ with $p_i$; this is usually called the \emph{propensity} of the observation.

We record a click log $\mathcal{L} = \{(R^{(j)}, c^{(j)})\}_{j=1}^{n}$, containing rankings of documents $R$ and clicks $c$ according to the procedure in Algorithm~\ref{alg:data}.
For brevity we drop the superscript notation $\cdot^{(j)}$ for the session identifier.
We now derive the learning objective of~\cite{joachims2017unbiased}, a modified version of the SVMRank training objective that minimizes the average rank of relevant results, weighted by the inverse propensity scores:
\begin{equation}
\hat{\theta} = \argmin_{\theta} \frac{1}{|\mathcal{L}|} \sum_{(R, c) \in \mathcal{L}} \sum_{\{i : c_i = 1\}} \frac{\text{rank}(R_i \mid f_\theta)}{p_i}.
\end{equation}
It can be shown that the above training objective is unbiased and can be solved via a hinge loss formulation~\cite{joachims2017unbiased}.
We will refer to this method as \acfi{CF-RANK}.
\label{sec:counterfactual:svm}

\begin{algorithm}[t]
	\caption{Data collection for \acl{CLTR}.} 
	\label{alg:data}
	\begin{algorithmic}[1]
		\STATE \textbf{Input}: production ranker: $f_{\text{production}}$; log size $n$;
		\STATE \textbf{Output}: a session log $\mathcal{L}$
		\STATE $\mathcal{L} = \emptyset$
		\FOR{$t \leftarrow  1 \ldots n$ }
		\STATE $q^{(t)} \leftarrow receive\_query(t)$
		\STATE $D^{(t)} \leftarrow preselect\_documents(q^{(t)})$
		\STATE $R^{(t)} \leftarrow rank\_documents(f_{\text{production}}, D^{(t)})$
		\STATE $c^{(t)} \leftarrow receive\_clicks(R^{(t)})$
		\STATE $\mathcal{L} \leftarrow \mathcal{L} \cup \{(R^{(t)}, c^{(t)})\}$
		\ENDFOR
	\end{algorithmic}
\end{algorithm}

\subsection{Unbiased LTR with additive metrics}
The counterfactual learning framework described in the previous section can be adapted to optimize additive metrics~\cite{agarwal2018counterfactual}.
For example, we can modify the training objective so it optimizes DCG~\cite{jarvelin2002cumulated}, a common metric in evaluation rankings: 
\begin{equation}
\hat{\theta} = \argmin_{\theta} \frac{1}{|\mathcal{L}|} \sum_{(R, c) \in \mathcal{L}} \sum_{\{i : c_i = 1\}} \frac{\lambda(\text{rank}(R_i \mid f_\theta))}{p_i},
\end{equation}
where $\lambda(r) = \frac{-1}{\log(1 + r)}$.
This objective is both continuous and sub-differentiable, making it possible to solve using existing gradient descent techniques.
We will refer to the DCG-optimizing counterfactual method as \acfi{CF-DCG}.
\label{sec:counterfactual:dcg}

The counterfactual \ac{LTR} algorithm is described in Algorithm~\ref{alg:cf}.
As input, the algorithm takes a set of weights (typically initialized to 0), a scoring function $f$, a learning rate $\mu$ and a click log $\mathcal{L}$.
The algorithm runs for a fixed number of epochs which trades off computation time for convergence.
The gradient is calculated on line~\ref{line:cf:gradient} where a clicked document is compared against every other document. 
The gradient is computed as a $\lambda$-modified hinge loss:
with $\lambda(r) = r$ this is the \ac{CF-RANK} method, which attempts to minimize the rank of relevant results;
and with $\lambda(r) = \frac{-1}{\log(1 + r)}$ we obtain the \ac{CF-DCG} method, which attempts to maximize DCG.
Finally, the ranking model is updated via stochastic gradient descent on line~\ref{line:cf:update}.

\subsection{Propensity estimation methods}
\label{sec:counterfactual:propensities}
Recent work in \ac{CLTR} has focused on estimating propensities from data~\cite{ai2018unbiased-sigir,wang2016learning,wang2018position}.
As the aim of our work is to compare counterfactual and online \ac{LTR} approaches, we consider propensity estimation beyond the scope of this paper and assume the propensity scores are known a priori.
This is a reasonable assumption, as practitioners typically first perform a randomization experiment to measure the observation probabilities before applying a counterfactual learning algorithm~\cite{bendersky2017learning}.

Our experimental setup allows us to measure the difference of counterfactual methods and online methods without confounding our results by the accuracy of the propensity estimator.

\begin{algorithm}[t]
	\caption{\acf{CLTR}.} 
	\label{alg:cf}
	\begin{algorithmic}[1]
		\STATE \textbf{Input}: initial weights: $\mathbf{\theta}$; scoring function: $f$; learning rate $\mu$; click log $\mathcal{L}$; number of epochs: $E$.
		\FOR{$e \leftarrow  1 \ldots E$ }
		\FOR{$(R, c) \in \mathcal{L}$}
		\FOR{$R_i : c_i = 1$}
		\STATE $\nabla f_{\theta} \leftarrow \mathbf{0}$ \hfill \textit{\small // initialize gradient} 
		\FOR{$R_j : R_j \neq R_i$}
		\STATE $\nabla f_{\theta} \leftarrow \nabla f_{\theta} + \nabla \left[ \lambda(hinge(f_{\theta}(R_i) - f_{\theta}(R_j)))) \right]$\\\hfill \textit{\small // (modified) hinge-loss gradient} \label{line:cf:gradient}
		\ENDFOR
		\STATE $\nabla f_{\theta} \leftarrow \frac{\nabla f_{\theta}}{p_i}$
		\STATE $\theta \leftarrow \theta + \mu \nabla f_{\theta_{t}}$ 
		\hfill \textit{\small // update the ranking model} \label{line:cf:update}
		\ENDFOR
		\ENDFOR
		\ENDFOR
	\end{algorithmic}
\end{algorithm}

\section{Online Learning to Rank}
\label{sec:oltr}

\acf{OLTR}~\cite{yue09:inter, hofmann_2013_reusing, Schuth2014a, Oosterhuis2018Unbiased} aims to learn by directly interacting with users.
\ac{OLTR} algorithms affect the data gathered during the learning process because they have control over what is displayed to users.
These interventions potentially allow for more efficient learning by requiring less user interactions to reach a certain level of performance.
Yet, an \ac{OLTR} algorithm has to simultaneously provide good results to the user and learn from their interactions with the displayed results~\cite{oosterhuis2017balancing}.
Thus besides unbiasedly learning from user interactions, the user experience during learning should also be maintained.
A great advantage of the online approach is that learned behavior is immediately applied.
However, this high level of responsiveness also means that an unreliable \ac{OLTR} method can decrease the user experience immediately, making it a potential risk.
Therefore, it is important for \ac{OLTR} methods to be reliable, i.e. unbiased and robust to noise.

In contrast to \ac{CLTR}, \ac{OLTR} methods do not explicitly model user behavior, i.e., they do not estimate observance probabilities.
Instead, they use stochasticity in the displayed results to handle selection and position biases.
In addition, their properties are only based on simple assumptions about user behavior~\cite{yue09:inter, Oosterhuis2018Unbiased}, e.g., \emph{a relevant document is more likely to be clicked than a non-relevant document}.
Thus, in cases where users are hard to model \ac{OLTR} may have an advantage over \ac{CLTR}.
In other areas of machine learning~\cite{settles2012active}, online (or active) approaches tend to be more efficient than algorithms without control over data gathering w.r.t. data requirements.
However, \ac{CLTR} and \ac{OLTR} methods have never been compared directly, thus currently we do not know if this advantage also generalizes to \ac{LTR} problems.

\subsection{Dueling bandit gradient descent}
\ac{DBGD}~\cite{yue09:inter} is the earliest \ac{OLTR} method and is based on interleaving: an unbiased online evaluation method.
Interleaving methods unbiasedly compare two ranking systems in the online setting~\cite{radlinski2013optimized, hofmann2011probabilistic, joachims03:evaluating}.
Therefore, interleaving can be used to recognize an improvement to a ranking system.
At each iteration \ac{DBGD} compares its current model with a sampled variation using interleaving.
If a preference towards the variation is inferred, the current model is updated in its direction.
Over time this process estimates a gradient descent and the model should oscillate towards an optimum w.r.t.\ user preference.
Most \ac{OLTR} methods published to date are extensions of \ac{DBGD}.
This includes, for instance, Multileave Gradient Descent~\cite{schuth2016mgd}, which compares multiple variations per iteration using multileaving~\cite{Schuth2014a, oosterhuis2017sensitive};
other methods reuse historical interactions to guide exploration~\cite{hofmann_2013_reusing, zhao2016constructing}.
All of these extensions improve on the initial learning speed of \ac{DBGD}. 
However, no extension has been shown to improve long-term performance~\cite{schuth2016mgd, oosterhuis2016probabilistic, Oosterhuis2018Unbiased}.
Moreover, even under ideal circumstances and with a very large number of iterations \ac{DBGD} is unable to reach levels of performance comparable to \ac{LTR} from labeled datasets \citep{Oosterhuis2018Unbiased, oosterhuis2017balancing}.
Despite these shortcomings, \ac{DBGD} is a key method in the field of \ac{OLTR}.

\subsection{Pairwise differentiable gradient descent}
\label{sec:PDGD}
In reaction to the shortcomings of \ac{DBGD}, recent work has introduced the \ac{PDGD} algorithm~\citep{Oosterhuis2018Unbiased}.
In contrast to \ac{DBGD}, \ac{PDGD} does not depend on sampling model variations or online evaluation methods.
Instead, \ac{PDGD} constructs a pairwise gradient at each interaction, from inferred user preferences between documents.
Algorithm~\ref{alg:pdgd} describes the \ac{PDGD} method in formal detail.
At each iteration, the algorithm waits until a query is issued by the user (line~\ref{line:pdgd:query}).
Then \ac{PDGD} creates a probability distribution over documents by applying a Plackett-Luce model with parameter $\tau \in \mathbb{R}_{>0}$ to the scoring function:
\begin{align}
P(d | D, \theta) = \frac{e^{\tau f_\theta(d)}}{\sum_{d' \in D} e^{\tau f_\theta(d')}}. \label{eq:pdgd:pl}
\end{align}
We introduce the $\tau$ parameter to control the \emph{sharpness} of the initial distribution, which indicates the confidence we have in the initial model.
Previous work only considered cold-start situations thus did not require this parameter~\cite{Oosterhuis2018Unbiased, Oosterhuis2019optimizing}.
From this distribution a result list is sampled (Line~\ref{line:pdgd:sample}) and displayed to the user.
Then \ac{PDGD} infers preferences between the clicked documents and the first unclicked document and every unclicked document preceding a clicked document, a longstanding pairwise assumption~\cite{joachims2002optimizing}.
With $d_i >_\mathbf{c} d_j$ denoting an inferred preference of $d_i$ over $d_j$, \ac{PDGD} estimates the model gradient as:
\begin{align}
\nabla f_\theta \approx \sum_{d_i >_\mathbf{c} d_j} \rho(d_i, d_j, R, \theta) \nabla P(d_i \succ d_j | \theta),
\end{align}
where $\rho$ is a weighing function used to deal with biases (line~\ref{line:pdgd:prefinfer}).
Finally, the scoring function is updated according to the estimated gradient (line~\ref{line:pdgd:update}), and the process repeats with the updated model.

The $\rho$ function makes use of the so-called reverse pair ranking function $R^*(d_i, d_j, R)$, which returns the same ranking as $R$ with the position of document $d_i$ and $d_j$ swapped.
Then, the value of $\rho$ is determined by the ratio between the probabilities of the two rankings:
\begin{align}
\rho(d_i, d_j, R, \theta) &=
\frac{
P(R^*(d_i, d_j, R) | \theta)
}{
P(R | \theta) + P(R^*(d_i, d_j, R) | \theta)
}.
\end{align}
\ac{PDGD} assumes that if a user considers both $d_i$ and $d_j$ equally relevant, then inferring the preference in $d_i >_\mathbf{c} d_j$ in $R$ is equally probable as inferring the reverse preference $d_j >_\mathbf{c} d_i$ in $R^*(d_i, d_j, R)$.
Furthermore, if a user prefers one of the documents, inferring the corresponding preference is more likely than the reverse.
These assumptions can formulated as:
\begin{align}
\begin{split}
\textit{sign}(&\textit{relevance}(d_i) - \textit{relevance}(d_j))
=
{}\\
&\textit{sign}\left(P(d_i \succ_\mathbf{c} d_j | R) - P(d_j \succ_\mathbf{c} d_i | R^*(d_i, d_j, R))\right).
\end{split}
\end{align}
Intuitively, this means that relative relevance differences can be inferred by swapping document pairs without changing the rest of the ranking.
A similar approach is used by counterfactual methods to estimate propensities~\cite{joachims2017unbiased}, conversely, \ac{PDGD} uses it to directly optimize its ranking model.
In the original paper \citeauthor{Oosterhuis2018Unbiased} prove that the gradient estimation of \ac{PDGD} is unbiased w.r.t. document pair preferences.
This means that the expected gradient of \ac{PDGD} can be written as a sum over all document pairs:
\begin{align}
E[\Delta f(\cdot, \theta)] = \sum_{(d_i, d_j) \in D} \alpha_{ij} \left(f'(d_i, \theta) - f'(d_j, \theta)\right),
\end{align}
where $\alpha_{ij}$ is a unique weight for every document pair in the collection.
\ac{PDGD} is unbiased in the sense that the sign of $\alpha_{ij}$ matches the user preferences between $d_i$ and $d_j$:
\begin{align}
\textit{sign}(\alpha_{ij}) = \textit{sign}\left(\textit{relevance}(d_i) - \textit{relevance}(d_j)\right).
\end{align}
Thus, in expectation \ac{PDGD} will perform an unbiased update towards the pairwise preferences of the user.

Recent work has extensively compared \ac{DBGD} with \ac{PDGD}~\cite{Oosterhuis2018Unbiased, Oosterhuis2019optimizing}; \ac{PDGD} performs considerably better in terms of final convergence, user experience during optimization, and learning speed. 
These findings generalize from settings with no to moderate levels of position bias and noise~\cite{Oosterhuis2018Unbiased} to circumstances with extreme levels of bias and noise~\cite{Oosterhuis2019optimizing}.
\ac{PDGD} is the new state-of-the-art for \ac{OLTR}, and we will therefore not consider \ac{DBGD} in our comparison.

\begin{algorithm}[t]
\caption{\acf{PDGD}.} 
\label{alg:pdgd}
\begin{algorithmic}[1]
\STATE \textbf{Input}: initial weights: $\mathbf{\theta}_1$; scoring function: $f$; learning rate $\mu$.
\FOR{$t \leftarrow  1 \ldots \infty$ }
	\STATE $q^{(t)} \leftarrow \textit{receive\_query}(t)$\hfill \textit{\small // obtain a query from a user} \label{line:pdgd:query}
	\STATE $D^{(t)} \leftarrow \textit{preselect\_documents}(q^{(t)} )$\hfill \textit{\small // preselect doc. for query} \label{line:pdgd:preselect}
	\STATE $\mathbf{R}^{(t)} \leftarrow \textit{sample\_list}(f_{\theta }, D^{(t)} )$ \hfill \textit{\small // sample list according to Eq.~\ref{eq:pdgd:pl}} \label{line:pdgd:sample}
	\STATE $\mathbf{c}^{(t)} \leftarrow \textit{receive\_clicks}(\mathbf{R}^{(t)} )$ \hfill \textit{\small // show result list to the user} \label{line:pdgd:clicks}
	\STATE $\nabla f_{\theta } \leftarrow \mathbf{0}$ \hfill \textit{\small // initialize gradient} \label{line:pdgd:initgrad}
	\FOR{$d_i >_{\mathbf{c}} d_j \in \mathbf{c}^{(t)}$} \label{line:pdgd:prefinfer}
	\STATE $\nabla f_{\theta } \leftarrow \nabla f_{\theta} + \rho(d_i, d_j, R) \nabla P(d_i \succ d_j | \theta )$
	\ENDFOR
	\STATE $\theta \leftarrow \theta  + \mu \nabla f_{\theta }$
	\hfill \textit{\small // update the ranking model} \label{line:pdgd:update}
\ENDFOR
\end{algorithmic}
\end{algorithm}

\section{Expectations from Previous Work}
\label{sec:hypotheses}

This section will discuss several expectations about the qualitative differences between \ac{CLTR} and \ac{OLTR} based on previous work.
Subsequently Section~\ref{sec:experiments} describes the experiments that have been run to test these expectations and Section~\ref{sec:results} their outcomes.
By discussing existing expectations here we can later contrast them with our observations.
Whether and how our results match our expectations can reveal how well our understanding of \ac{LTR} from user interactions is.

\myparagraph{Expectation 1 -- The performance at convergence}
As described in Section~\ref{sec:counterfactual} it has been proven that \ac{CLTR} can unbiasedly optimize additive metrics~\cite{agarwal2018counterfactual}, for instance using \ac{CF-DCG}, when the observation probabilities of the user are correctly known.
Conversely, for \ac{PDGD} there is no known proof that it optimizes any metric unbiasedly.
Therefore, we expect \ac{CLTR} methods like \ac{CF-DCG} to reach a higher level of performance than \ac{PDGD} if the propensities are known, since \ac{CLTR} can guarantee that the performance metric is optimized, while for \ac{PDGD} it is unclear whether its pairwise gradient will optimize the metric precisely.

\myparagraph{Expectation 2 -- The user experience during learning}
The field of \ac{OLTR} has long claimed that their methods provide the most responsive experience~\cite{schuth2016mgd, hofmann_2013_reusing, Oosterhuis2018Unbiased} because \ac{OLTR} methods apply their learned model instantly.
However, noise may cause a method to decrease model quality (temporarily) and exploration adds stochasticity to the results, thus risking a worsened user experience.
As a result, we expect an \ac{OLTR} method to provide an experience worse than the initial ranker at the start, but as learning continues the user experience should eventually exceed that of the initial model.
In contrast, \ac{CLTR} methods do not affect the user experience during learning as they work with historical data, and therefore, also cannot improve it.
Nevertheless, this approach completely avoids the risks of degrading the user experience.
Therefore, we expect \ac{OLTR} to provide a worse experience than under click gathering for \ac{CLTR} initially, yet eventually the experience under \ac{OLTR} should exceed that of \ac{CLTR}.
The question is whether the long-term improvements of \ac{OLTR} outweigh the initial decrease.

\myparagraph{Expectation 3 -- The effect of interventions}
Interventions are expected to greatly reduce the data requirements for learning~\cite{settles2012active}, as they allow algorithms to gather data that is more useful for their current state.
Correspondingly, \ac{OLTR} methods are expected to learn faster~\cite{hofmann_2013_reusing, schuth2016mgd}, in other words, they should require less user interactions to reach a decent level of performance than \ac{CLTR} methods~\cite{Oosterhuis2018Unbiased}.
Similarly, allowing \ac{CLTR} methods to intervene, e.g., by deploying a current model should make them more efficient as well.

\smallskip\noindent%
This concludes the key expectations regarding the performance differences between \ac{CLTR} and \ac{OLTR} methods.
While these expectations are based on previously published literature on \ac{CLTR} and \ac{OLTR}~\cite{Oosterhuis2018Unbiased, settles2012active, agarwal2018counterfactual}, they have never directly been tested.
To the best of our knowledge, our study is the first to confirm or challenge them with hard experimental facts.


\begin{table}[tb]
\caption{Click probabilities after observing a document in the result list for different user models.}
\centering
\begin{tabular}{ l c c c c c }
\toprule
& \multicolumn{5}{c}{ $P(\text{click}=1 | \text{observed}=1, \text{rel}(d))$} \\
\midrule
$\text{rel}(d)$ & \emph{$ 0$} & \emph{$ 1$}  &  \emph{$ 2$} & \emph{$ 3$} & \emph{$ 4$} \\
\midrule
 \emph{Perfect} &  0.00 &  0.20 &  0.40 &  0.80 &  1.00  \\
 \emph{Binarized} &  0.10 &  0.10 &  0.10 &  1.00 &  1.00  \\
 \emph{Near-Random} &  0.40 &  0.45 &  0.50 &  0.55 &  0.60  \\
\bottomrule
\end{tabular}
\label{tab:clickmodels}
\end{table}

\section{Experiments}
\label{sec:experiments}

Our experiments evaluate the user experience of several methods at different time-steps and a multitude of conditions with varying levels of interaction noise, position bias, and selection bias.
Due to the scope of this comparison and the varying requirements, we rely on a synthetic setting based on an existing \ac{LTR} dataset and simulated user behavior.
Our setup is an extension of the synthetic experiments common in both \ac{OLTR}~\cite{hofmann_2013_reusing, schuth2016mgd, Oosterhuis2018Unbiased} and \ac{CLTR}~\cite{joachims2017unbiased, ai2018unbiased-sigir}.

\subsection{Optimization setup}
\label{sec:experiments:setup}

We use the \emph{Yahoo! Webscope} dataset~\cite{chapelle2011yahoo}; it contains a set of queries with a unique set of preselected documents for each query.
The dataset provides a train, validation and test split. We use the train partition during optimization of the methods, the validation set for tuning hyperparameters and the test partition to report our results.
Each query-document pair is represented by a feature vector and a relevance label, the relevance labels are in a five-degree scale ranging from \emph{not relevant} (0) to \emph{perfectly relevant} (4).

A baseline ranker is trained to serve as a logging ranker for the \ac{CLTR} methods, and an initial ranker to warm-start the \ac{OLTR} method.
To create the baseline ranker, we follow the setup of~\citet{joachims2017unbiased} and train an SVMRank ranker on 1\% of the queries in the training data.
This setup is chosen as it reflects a common real-world scenario: it is possible to manually annotate a small amount of data to learn an initial ranker, and then use a large amount of logged interaction data, either online or counterfactually, to further improve this ranker.

Finally, the gathering of click-data is simulated using the following steps:
First, a user-issued query is simulated by uniformly sampling a query from the training partition of the dataset.
Then, the corresponding documents are ranked according to the applied \ac{LTR} method, i.e., by the logging policy for \ac{CLTR} methods or by the algorithm itself in \ac{OLTR}.
Subsequently, the ranked results are displayed to a simulated user who then clicks on any number of documents (including none); Section~\ref{sec:experiments:behavior} details the behavior models we applied.
Lastly, the resulting clicks are presented to the \ac{LTR} method, which may now use the interaction for optimization.

\subsection{Simulating user behavior}
\label{sec:experiments:behavior}

We simulate user behavior by modelling three aspects of user behavior in search: interaction noise, position bias and selection bias.

First, \emph{interaction noise} affects the probability of a user clicking on a document after observing it.
The probability of clicking is conditioned on the relevance label of a document, as users are more likely to click on more relevant documents.
Table~\ref{tab:clickmodels} provides the click probabilities for three different click behavior models:
\emph{Perfect} click behavior has probabilities proportional to the relevance and never clicks on a non-relevant document, simulating an \emph{ideal} user.
\emph{Binarized} click behavior acts on only two levels of relevance and is affected by position-bias; this simulated behavior has been used in previous work on \ac{CLTR}~\cite{joachims2017unbiased,agarwal2018counterfactual,ai2018unbiased-sigir}.
And \emph{Near-Random} behavior clicks very often, and only slightly more frequently on more relevant documents than on less relevant documents; this behavior simulates very high levels of click noise.

Second, \emph{position bias} is modelled by observation probabilities; for a document at rank $i$ the probability of being observed is determined by the parameter $\eta$ and formula:
\begin{align}
P(observed = 1 \mid i) = \left(\frac{1}{i}\right)^\eta.
\end{align}
Again this follows previous work on \ac{CLTR}~\cite{joachims2017unbiased,agarwal2018counterfactual,ai2018unbiased-sigir}.
We apply this position bias to the \emph{Binarized} and \emph{Near-Random}  user models; the \emph{Perfect} user observes all documents every time and thus has no position bias.
In our experiments we use $\eta=1$ and $\eta=2$ to model different levels of position bias.

Thirdly, we simulate \emph{selection bias}, which occurs when not all documents can be displayed and thus also not observed.
In practice it also occurs because users never look past certain positions, for instance, users rarely look beyond the initial page of many multi-page result displays.
We model selection bias by giving a zero observance probability to documents beyond rank $10$. 
This is common practice in \ac{OLTR} experiments~\cite{hofmann_2013_reusing, schuth2016mgd, Oosterhuis2018Unbiased}; in contrast, \ac{CLTR} methods assume that no selection bias is present.
To investigate the effect of selection bias, our experiments both contain simulations with and without it.

In conclusion, we can apply selection bias, have two levels of position bias, and three levels of interaction noise.
In total, we apply ten different types of user behavior: \emph{Perfect} click behavior with and without selection bias, the \emph{Binarized} and \emph{Near-Random} click behaviors with two levels of position bias, with and without selection bias.
To the best of our knowledge this is the most extensive set of types of behavior used for evaluating \ac{CLTR} and \ac{OLTR} methods, in addition to being the first comparison between the two methodologies.

\subsection{Evaluation}
\label{sec:experiments:metrics}
To measure the performance of a ranker at any time step, we evaluate it on held-out annotated test data using the $nDCG@10$ metric~\cite{jarvelin2002cumulated}.
We use the learned models without any exploration when evaluating performance at convergence.
To evaluate the user experience during learning for \ac{OLTR} we apply the algorithm with exploration to the held-out dataset, this measures the performance for a previously unseen query that appears during optimization.

To test for statistical significance we average the results over multiple runs and apply a two-tailed t-test.
In Section~\ref{sec:results}, we indicate for each comparison, whether the observed difference is statistically significant with $p < 0.01$.

\subsection{Methods and interventions}
\label{sec:experiments:methods}

Our comparisons concern one \ac{OLTR} method and several \ac{CLTR} methods, in addition to \ac{CLTR} methods with periodic deployments during the optimization process.

The \ac{OLTR} method in the comparison is \ac{PDGD}~(Section~\ref{sec:oltr}); the parameters ($\mu = 0.01$, $\tau=10$) were tuned on the validation set.

The \ac{CLTR} methods we apply are \ac{CF-RANK} (Section~\ref{sec:counterfactual:svm}) and \ac{CF-DCG} (Section~\ref{sec:counterfactual:dcg}); the former is the original \ac{CLTR} method while the latter optimizes DCG corresponding to our evaluation metric.
For each run, the \ac{CLTR} methods are given the propensity scores of the actual user models applied; this guarantees that the \ac{CLTR} methods optimize unbiasedly.
Furthermore, to investigate the effect of interventions we also run these methods with \emph{periodic deployment}.
For these runs we replaced the logging policy with the (then) current ranking model after every 200,000 iterations, thus simulating a situation where the learned model is deployed periodically.
The parameters for the \ac{CLTR} were optimized for every instance of user behavior and number of interactions on the validation set, thus results at different time-steps may use different parameter settings.
The complete set of hyper parameter settings that were used is released alongside our source code; see Section~\ref{sec:conclusion} for details.


\begin{figure*}[tb]
	\vspace{-0.2cm}
	\includegraphics[width=0.99\textwidth]{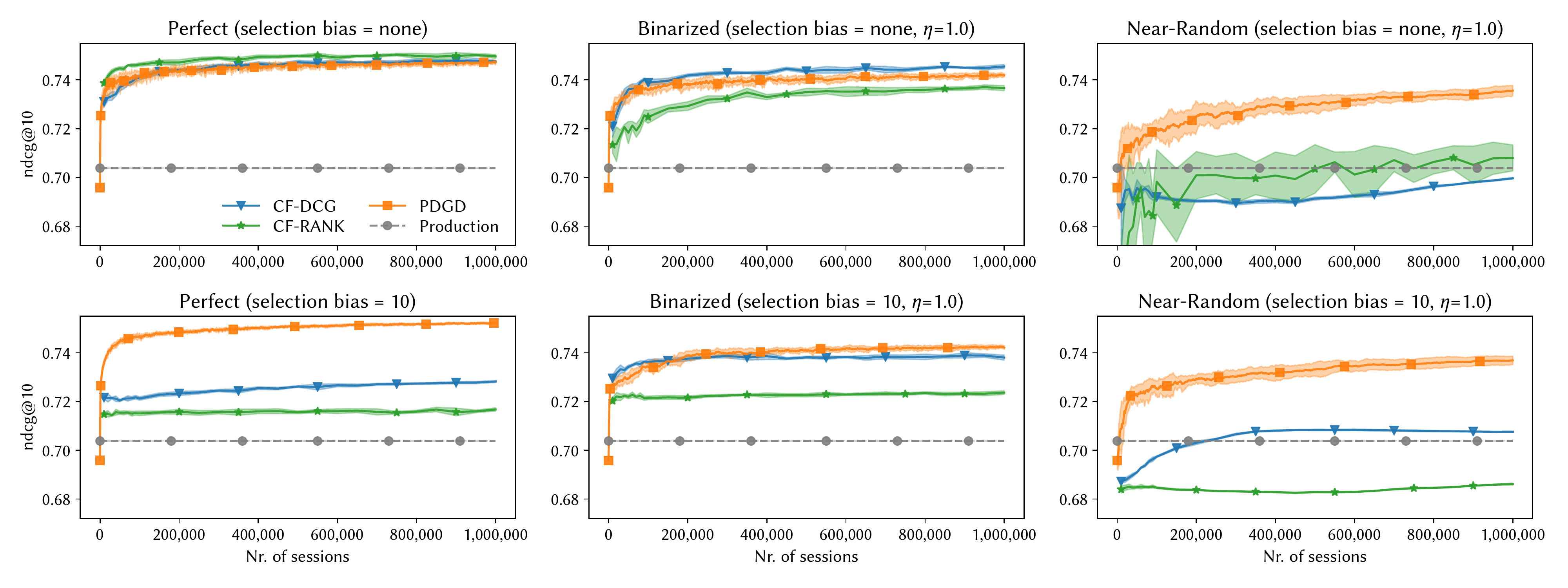}	
	\vspace{-0.2cm}
	\caption{Performance of online and counterfactual methods under perfect, binarized, and near-random user models. In the top row no selection bias is present; in the bottom row, a selection of 10 is used.}
	\label{fig:performance}
\end{figure*}

\begin{figure*}[tb]
	\vspace{-0.2cm}
	\includegraphics[width=0.67\textwidth]{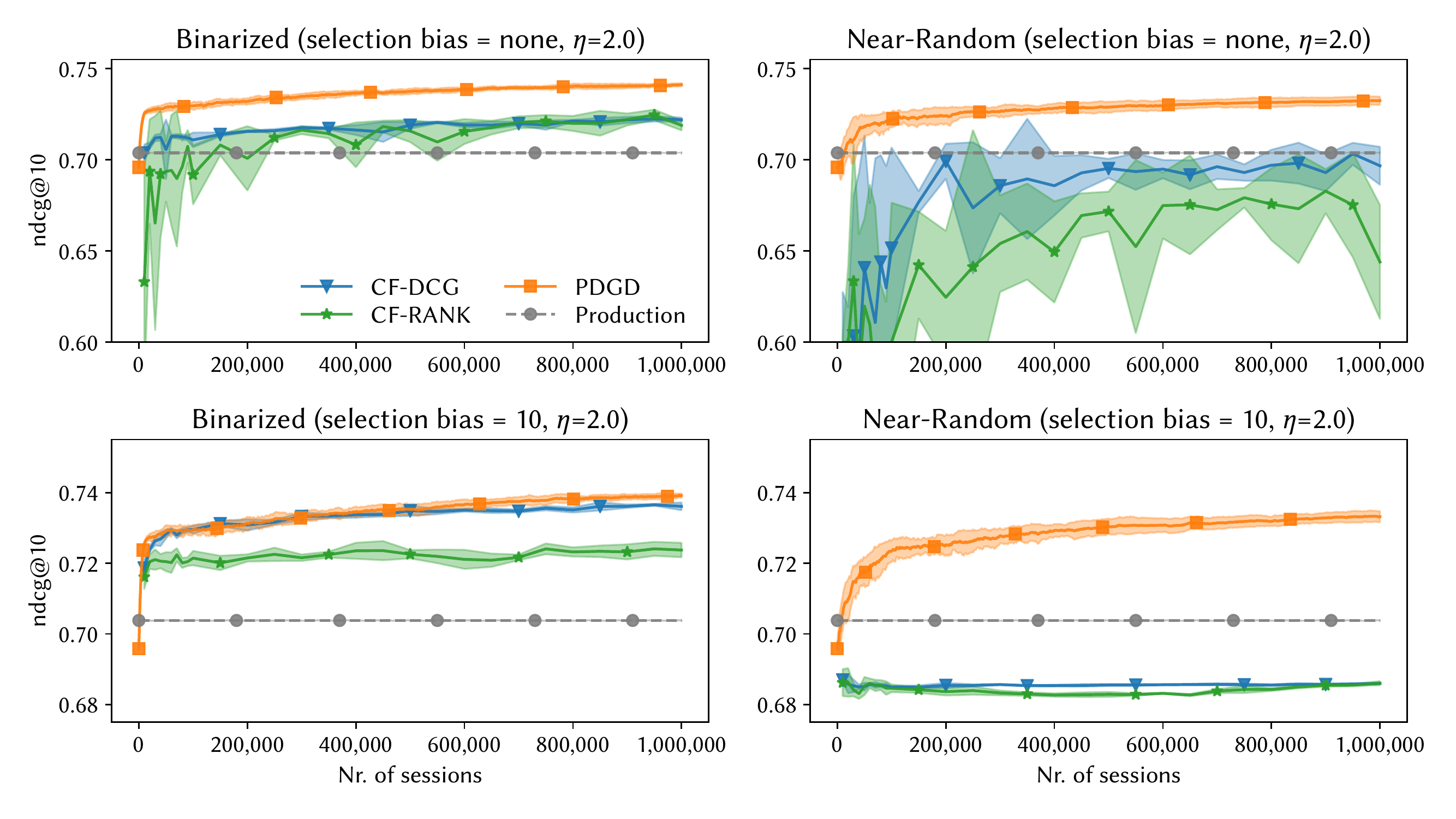}	
	 \vspace{-0.2cm}
	\caption{Performance of online and counterfactual methods under very strong position bias ($\eta=2$). The scale of the y-axis of the plots in the top row has been modified to be able to show the large variance. In the top row no selection bias is present; in the bottom row, a selection of 10 is used.}		
	\label{fig:heavybias}
\end{figure*}

\begin{figure*}
	\vspace{-0.2cm}
	\includegraphics[width=0.99\textwidth]{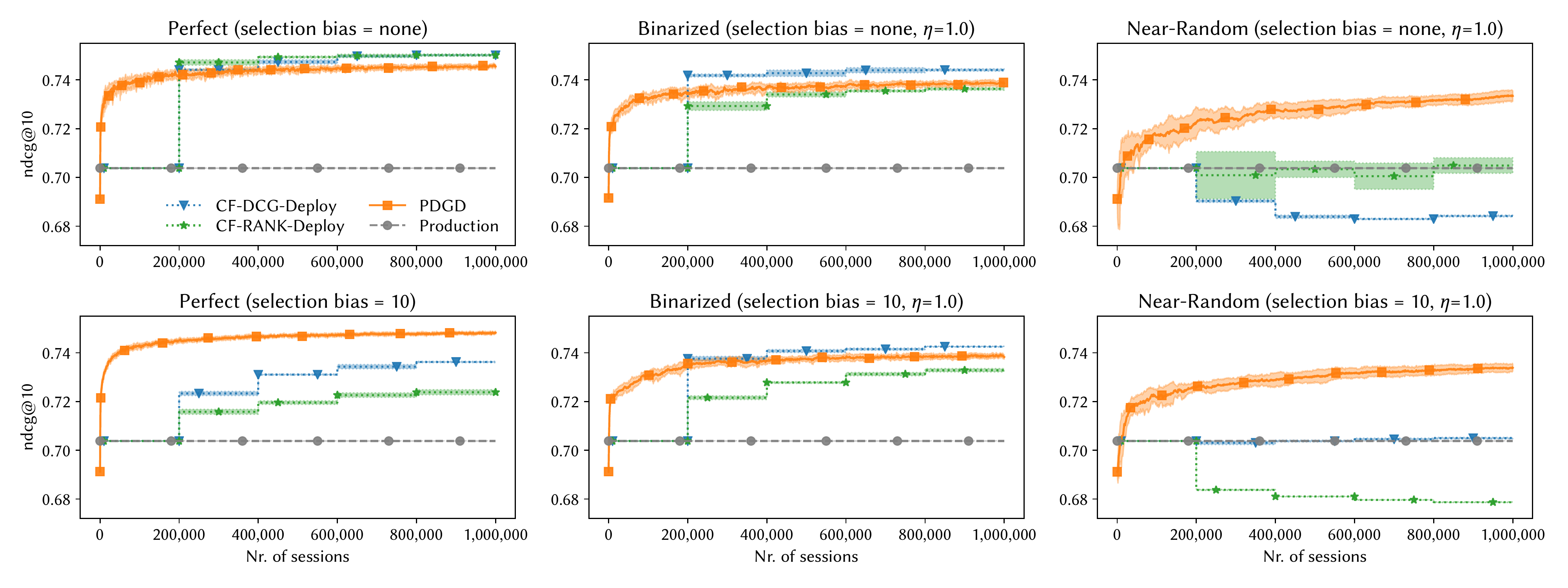}
	\vspace{-0.2cm}
	\caption{Display performance during training, indicating user experience. In the top row no selection bias is present; in the bottom row, a selection of 10 is used.}		
	\label{fig:display}
\end{figure*}

\begin{figure*}
	\vspace{-0.2cm}
	\includegraphics[width=0.99\textwidth]{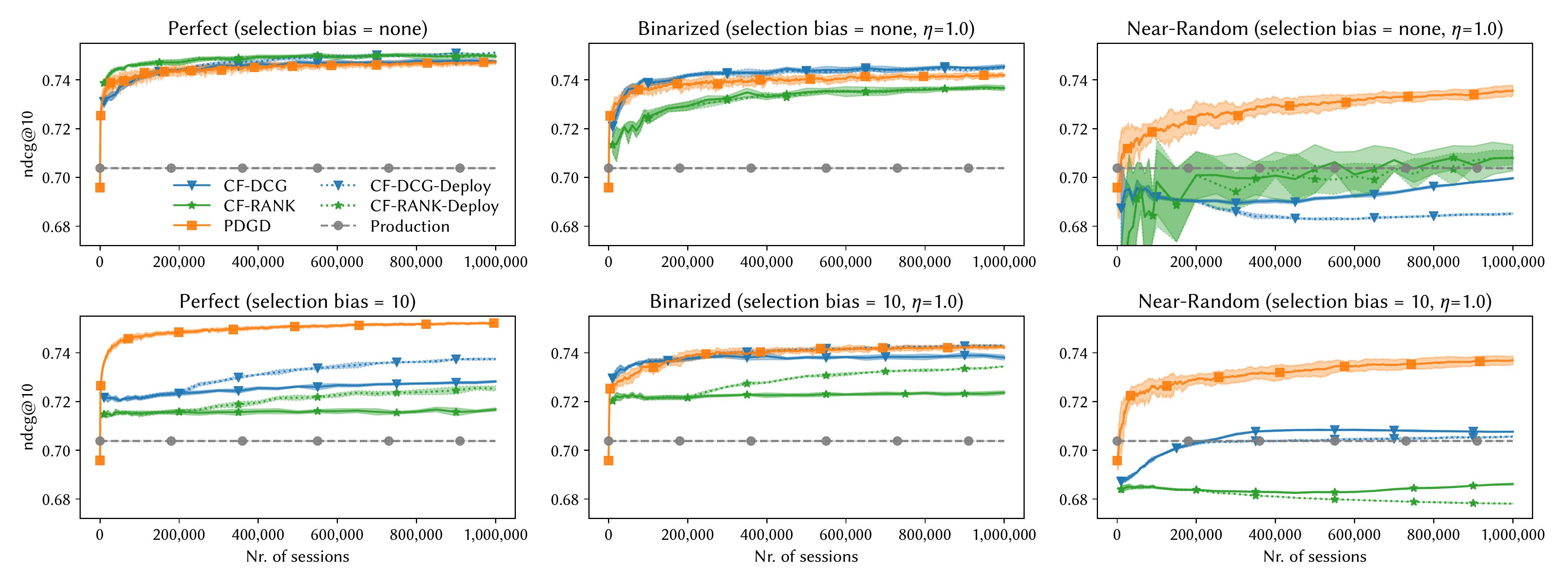}	
	\vspace{-0.2cm}
	\caption{Performance of counterfactual methods with a deployment every 200,000 sessions. In the top row no selection bias is present; in the bottom row, a selection of 10 is used.}
	\label{fig:intervention}
\end{figure*}

\section{Results and Analysis}
\label{sec:results}
This section will answer the research questions posed in Section~\ref{sec:intro}, using our experimental results displayed in Figure~\ref{fig:performance},~\ref{fig:heavybias},~\ref{fig:display},~and~\ref{fig:intervention}.

\subsection{Ranking performance}

We investigate the ranking performance of the \ac{OLTR} and \ac{CLTR} methods under different experimental conditions to answer \ref{rq:performance}:
\begin{itemize}
\item[] \emph{Do state-of-the-art online and counterfactual \ac{LTR} \mbox{methods} converge at the same level of performance?}
\end{itemize}
As discussed in Section~\ref{sec:hypotheses} we expect \ac{CF-DCG} to have better performance when the assumptions for unbiased \ac{CLTR} are met, as unlike \ac{PDGD}, \ac{CF-DCG} is then proven to optimize the DCG metric.
Figure~\ref{fig:performance} displays the performance of the counterfactual and online \ac{LTR} methods over 1,000,000 sessions, where a session consists of a query, a ranking of documents and all corresponding clicks generated by the click simulation.
These results are grouped by the user behavior under which they were optimized, which varies in the amount of selection bias, position bias, and interaction noise.

First, we consider the results without selection bias displayed in the top row of Figure~\ref{fig:performance}.
Under \emph{Perfect} and \emph{Binarized} click behavior the performance of the \ac{CLTR} methods and \ac{PDGD} are quite comparable, with \ac{CF-DCG} performing better than \ac{PDGD} in the \emph{Binarized} case ($p < 0.01$).
In contrast, performance of the \ac{CLTR} methods drops below the production ranker under \emph{Near-Random} click behavior, and does not exceed it within 1,000,000 iterations ($p < 0.01$).
This goes against our expectations, as the \emph{Near-Random} case does not violate any assumptions necessary for unbiased \ac{CLTR} as described in~\cite{joachims2017unbiased}.
\ac{PDGD}, on the other hand, is much less affected and reaches much higher levels of performance.
Because the \emph{Binarized} and \emph{Near-Random} behaviors have the same position bias, the difference in performance must be due to the increased interaction noise in the latter.
Thus, it appears that the \ac{CLTR} methods are much less robust to noise than \ac{PDGD}, yet with low levels of noise \ac{CLTR} methods outperform the \ac{OLTR} method.

Second, we look at the results with selection bias displayed in the bottom row in Figure~\ref{fig:performance}.
For each user model the performance of the \ac{CLTR} methods is worse than without selection bias.
Except under \emph{Near-Random} click behavior where \ac{CF-DCG} now performs slightly better than the production ranker.
Unbiased \ac{CLTR} does not consider selection bias~\cite{joachims2017unbiased} which could explain this unexpected result.
In contrast, the performance of \ac{PDGD} is affected very little in comparison and is now better than both \ac{CLTR} methods under all click behaviors ($p < 0.01$).
Thus, it appears that \ac{PDGD} is preferable when selection bias is present.

Third, to understand the effect of position bias we look at the results in Figure~\ref{fig:heavybias}, where strong position bias is simulated with $\eta=2$.
It is clear that all methods are negatively affected by strong position bias.
Unexpectedly, \ac{PDGD} now outperforms the \ac{CLTR} methods in all cases ($p < 0.01$), even though the \emph{Binarized} click behavior without selection bias provides the exact circumstances for which \ac{CLTR} was designed~\cite{joachims2017unbiased}.
Therefore, we attribute the negative effect on \ac{CLTR} to high variance since the methods are still proven to be unbiased in this case.
This may further explain why selection bias has a positive effect on the \ac{CLTR} methods under the \emph{Binarized} click behavior: it removes documents with low propensities that lead to high variance.
Clearly, we see that \ac{OLTR} is better at handling high levels of position bias than \ac{CLTR}.

In conclusion, we answer \ref{rq:performance} negatively: online and counterfactual methods do not converge to the same level of performance.
However, which method reaches the best performance depends heavily on the conditions under which they are deployed.
In the presence of selection bias or under high levels of position bias or interaction noise \ac{OLTR} reaches the highest performance.
However, when there is no selection bias, little position bias and little interaction noise, then \ac{CLTR} reaches a level of performance that \ac{OLTR} is unable to obtain.
Counter to our expectations, even when the assumptions of \ac{CLTR} are true, the \ac{CLTR} methods are still not robust to noise.
Thus, to be able to make the best decision between the \ac{CLTR} and \ac{OLTR} methodologies, a practitioner should first measure the severity of different types of bias and noise in their search scenario.

\subsection{User experience}
In this section, we examine the quality of displayed rankings in order to answer \ref{rq:userexperience}:
\begin{itemize}
\item[] \emph{Is the user experience the same for online methods as for counterfactual methods?}
\end{itemize}
Figure~\ref{fig:display} shows the quality of rankings displayed by the \ac{PDGD} method during optimization and of the \emph{Production} ranker used to gather click-logs for the \ac{CLTR} methods.
For clarity: we are not discussing the \emph{CF-DCG-Deploy} and \emph{CF-RANK-Deploy} results for this research question, they will be discussed in Section~\ref{sec:results:interventions}.

In Section~\ref{sec:hypotheses} we stated the expectation that \ac{OLTR} methods start with a user experience worse than the production ranker due to exploration.
However, \ac{OLTR} is expected to overtake the production ranker as it continues to learn from interactions.
The results in Figure~\ref{fig:display} confirm this expectation.
Across all types of user behavior, we see that the initially displayed performance is substantially worse than the production ranker ($p < 0.01$).
\ac{PDGD} provides considerably better rankings than the production ranker within 1,000, 2,000 and 21,000 sessions for \emph{Perfect}, \emph{Binarized} and \emph{Near-Random} click behavior, respectively ($p < 0.01$).
Thus, we conclude that \ac{PDGD} provides a better user experience than \ac{CLTR} methods overall, with a decrease in quality for a limited initial period.

Therefore, we answer \ref{rq:userexperience} negatively: \ac{OLTR} does not provide the same user experience as \ac{CLTR}.
Besides a limited initial period, \ac{OLTR} provides a more responsive user experience during optimization than \ac{CLTR}.
However, it is up to practitioners to decide whether the initial worse period is worth it, or whether they prefer the constant user experience in the click gathering for \ac{CLTR}.

\subsection{The power of interventions}
\label{sec:results:interventions}

In this section we investigate whether performing interventions helps the learning performance, so as to answer \ref{rq:interventions}:
\begin{itemize}
\item[] \emph{When do online interventions help the learning \mbox{algorithm}?}
\end{itemize}
To answer this question we consider the performance of the optimized models in Figure~\ref{fig:intervention}, and the user experience during click gathering in Figure~\ref{fig:display}.

In Section~\ref{sec:hypotheses} we stated the expectation that interventions significantly speed up the learning process.
In Figure~\ref{fig:intervention} the performance of \ac{CLTR} methods diverge at the first moment of deployment: after 200,000 sessions.
We see that only in cases with high interaction noise, i.e., \emph{Near-Random} click behavior, periodic deployment leads to worse performance than without ($p < 0.01$).
For \emph{Perfect} and \emph{Binarized} click behavior, periodic deployment has no negative effects, moreover, when selection bias is present it substantially increases performance ($p < 0.01$).
Thus it appears that interventions cause \ac{CLTR} methods to reach higher levels of performance and especially help in dealing with selection bias.

Then we examine Figure~\ref{fig:display}, which displays the user experience during click gathering.
Here we see that interventions allow users to benefit from improvements earlier, or suffer from deteriorations sooner.
The same trend appears: a worse experience under high interaction noise, a better experience with little noise.
Furthermore, \ac{CF-DCG} with periodic deployment is capable of providing a better user experience than \ac{PDGD} when little noise is present ($p < 0.01$).
Unlike \ac{PDGD}, \ac{CF-DCG} does not perform exploration which seems to be a crucial advantage in these cases.

Lastly, we discuss the expectation that interventions speed up learning, in particular that \ac{OLTR} methods require significantly less data.
None of our results indicate that \ac{OLTR} learns faster than \ac{CLTR} methods.
While in many cases \ac{OLTR} reaches higher performance levels than \ac{CLTR}, when they reach comparable levels they do so after similar numbers of interactions.
We suspect the reason to be that \ac{PDGD} does not reiterate over previous interactions, where the \ac{CLTR} methods perform numerous epochs.
Nonetheless, despite expectations in previous work our results do not indicate that the interventions of \ac{OLTR} reduce data requirements.

To answer \ref{rq:interventions}: interventions help \ac{CLTR} methods in circumstances where they already improve over the production ranker.
Moreover, their effect is substantial when dealing with selection bias.
Unfortunately, deployment in difficult circumstances, i.e. with high levels of noise, can decrease performance even further and negatively affect the user experience considerably.
Thus, practitioners should realize that a repeated cycle of optimization and deployment with \ac{CLTR} can be quite harmful to the user experience.
Counterfactual evaluation~\citep{li2011unbiased,thomas2015high,swaminathan2017off} could estimate whether the deployment of a model improves the experience, before deployment.
The question is whether this evaluation is reliable and sensitive enough to prevent harmful changes.


\section{Related Work}
\label{sec:related}

In this section we discuss previous work that concerns large-scale comparisons of \ac{LTR} methods. 

\citet{liu2009learning} provides a comprehensive overview of the (then) state-of-the-art in (mostly) \ac{LTR} from labeled data but does not include a large-scale empirical comparison of methods.
\citet{tax2015cross} do compare \ac{LTR} algorithms that use manually annotated training data in a large-scale cross-benchmark comparison.
They show that there is no single optimal \ac{LTR} algorithm and provide a selection of supervised \ac{LTR} methods that are pareto-optimal.
In this paper we compare two different families of \ac{LTR} algorithms: \emph{online} and \emph{counterfactual} \ac{LTR} methods, neither of which learn from manually annotated data; both types of method utilize user interactions.
As such, the algorithms we compare are not supervised in the traditional sense~\cite{joachims2017unbiased}.

A systematic comparison of \ac{CLTR} methods appears to be lacking at this moment in time.
\citet{joachims-2016-counterfactual} seem to have provided the first comprehensive overview of counterfactual methods for \ac{LTR} aimed at the information retrieval community, but the authors do not include a large-scale experimental comparison.
More recently, \citet{ai2018unbiased-cikm} provide an overview of existing approaches to \ac{CLTR} and describe both the theory and detailed instructions on how to deploy \ac{CLTR} in practice.
Furthermore, their work also contrasts \ac{CLTR} with click models~\cite{chuklin2015click} but it does not contrast \ac{CLTR} and \ac{OLTR} methods.

Similarly, a systematic comparison of \ac{OLTR} methods appears to be lacking too.
The comprehensive survey due to \cite{grotov2016online} is several years old; it does not provide a large-scale experimental comparison nor does it contrast \ac{CLTR} and \ac{OLTR} methods; modern \ac{OLTR} algorithms such as \ac{PDGD} are also not included. 
In a more recent tutorial on \ac{OLTR}, \citet{oosterhuis-2018-online} does provide a theoretical and experimental comparison of \ac{OLTR} methods based on \acl{DBGD} and \ac{PDGD}.

Our aim in this study is to gain an understanding in what situations counterfactual and online \ac{LTR} approaches are appropriate to be used.
To the best of our knowledge, there is no prior work that systematically compares counterfactual and  online \ac{LTR} approaches, or answers this question.


\section{Conclusion}
\label{sec:conclusion}

The goal of this study was to answer the question:
\begin{itemize}\em
\item[] How should \ac{LTR} practitioners choose which method to apply from either \emph{counterfactual} or \emph{online} \ac{LTR} methodologies?
\end{itemize}
The choice between \ac{OLTR} and \ac{CLTR} is important as there are large differences between the results obtained by the two methodologies.
We recognize three factors that determine which approach should be preferred: selection bias, position bias, and interaction noise.
\ac{CLTR} reaches a higher level of performance than \ac{OLTR} in the absence of selection bias, and when there is little position bias or interaction noise.
In contrast, \ac{OLTR} outperforms \ac{CLTR} in the presence of selection bias, high position bias, high interaction noise, or any combination of these three.
Surprisingly, \ac{CLTR} methods can decrease performance w.r.t. the production ranker when high levels of noise are present, even in situations where they are proven to be unbiased.
We conclude that \ac{OLTR} is more robust to different types of bias and noise than \ac{CLTR}.
Therefore, practitioners should be well aware of the levels of bias and noise present in their search setting to choose between the two methodologies.

Unlike \ac{OLTR}, \ac{CLTR} does not need to intervene and can use data collected by the production ranker, which prevents a negative impact on the user experience. 
\ac{OLTR} initially provides an experience worse than the production ranker but very quickly substantially improves over it.
We have not observed \ac{OLTR} having large negative effects on the user experience, even under high levels of interaction noise.
However, practitioners should consider whether they are willing to risk the initial worsened user experience in order to get long term gains.

We observed that cycles of optimization and deployment with \ac{CLTR} methods can have harmful effects on performance and user experience.
High levels of interaction noise can severely worsen model quality for \ac{CLTR}; if, subsequently, such a model is deployed, it can worsen the next model even further.
Thus, practitioners should realize that \ac{CLTR} brings risks to the user experience and evaluate models before deploying them, for instance using offline or counterfactual evaluation~\citep{li2011unbiased,thomas2015high,swaminathan2017off} .

Our comparison is not without limitations:
In our experiments, the \ac{CLTR} methods were provided with the exact propensities; in realistic settings these values are not known and have to be estimated~\citep{ai2018unbiased-sigir}.
Thus we do not consider how errors in propensity estimation affect the comparison.
Additionally, we have not applied any heuristic methods such as propensity clipping; these methods reduce variance but make \ac{CLTR} biased.
Such heuristics could help in situations with high position bias, though they would not help combat interaction noise.
Finally, our comparative study considers only a single metric on a single dataset. Although the dataset and metric we use are widely accepted as a benchmark in both \ac{OLTR} and \ac{CLTR}, we would like to extend our study to multiple datasets, measuring across various dimensions and utilizing real user behavior from deployed systems in future work.

Our findings also reveal the importance of safety in \ac{LTR}. Future work could further look at methods that evaluate when it is safe to deploy models.
Moreover, a method that could estimate which \ac{CLTR} or \ac{OLTR} algorithm would perform best, could provide the best user experience with little effort from practitioners.

\subsection*{Code and data}
To facilitate reproducibility of the results in this paper, we are sharing all resources used in this paper at \url{http://github.com/rjagerman/sigir2019-user-interactions}.

\begin{acks}
This research was partially supported by
Ahold Delhaize,
the Association of Universities in the Netherlands (VSNU),
the Innovation Center for Artificial Intelligence (ICAI),
and
the Netherlands Organisation for Scientific Research (NWO)
under pro\-ject nr
612.\-001.\-551.
All content represents the opinion of the authors, which is not necessarily shared or endorsed by their respective employers and/or sponsors.
\end{acks}

\bibliographystyle{ACM-Reference-Format}
\bibliography{sigir2019-cf-v-o}

\end{document}